\documentclass[11pt,titlepage,leqno]{article}

\usepackage{latexsym}
\usepackage{amsfonts}
\usepackage{amsmath}
\usepackage{amssymb}

\makeatletter
\newcount\hour  \newcount\minutes  \hour=\time  \divide\hour by 60
\minutes=\hour  \multiply\minutes by -60  \advance\minutes by \time
\def\mmmddyyyy{\ifcase\month\or Jan\or Feb\or Mar\or Apr\or May\or Jun\or Jul\or
  Aug\or Sep\or Oct\or Nov\or Dec\fi \space\number\day, \number\year}
\def\hhmm{\ifnum\hour<10 0\fi\number\hour :%
  \ifnum\minutes<10 0\fi\number\minutes}

\newlength{\filength}
\newsavebox{\gcbox}
\sbox{\gcbox}{\framebox[\filength]{\rule{0ex}{2ex}}}

  \newtheorem{theorem}{Theorem}[section]
  \newtheorem{corollary}[theorem]{Corollary}
  \newtheorem{claim}[theorem]{Claim}

\newcommand{\qedblob}{\mbox{\rule[-1.5pt]{5pt}{10.5pt}}}
\def\literalqed{{\ \nolinebreak\hfill\mbox{\qedblob\quad}}}

\def\qed{\literalqed}

  \newtheorem{goal}[theorem]{Goal}
  \newtheorem{lemma}[theorem]{Lemma}

  \newtheorem{definition}[theorem]{Definition}
\makeatletter
\def\@citex[#1]#2{\if@filesw\immediate\write\@auxout{\string\citation{#2}}\fi
  \def\@citea{}\@cite{\@for\@citeb:=#2\do
    {\@citea\def\@citea{,\linebreak[0]}\@ifundefined
       {b@\@citeb}{{\bf ?}\@warning
       {Citation `\@citeb' on page \thepage \space undefined}}%
\hbox{\csname b@\@citeb\endcsname}}}{#1}}
\newcommand{\singlespacing}{\let\CS=
\@currsize\renewcommand{\baselinestretch}{1}\tiny\CS}
\newcommand{\singlespacingplus}{\let\CS=
\@currsize\renewcommand{\baselinestretch}{1.25}\tiny\CS}
\newcommand{\doublespacing}{\let\CS=
\@currsize\renewcommand{\baselinestretch}{1.75}\tiny\CS}
\newcommand{\extradoublespacing}{\let\CS=
\@currsize\renewcommand{\baselinestretch}{1.9}\tiny\CS}
\newcommand{\draftspacing}{\let\CS=
\@currsize\renewcommand{\baselinestretch}{2.0}\tiny\CS}
\newcommand{\hugedraftspacing}{\let\CS=
\@currsize\renewcommand{\baselinestretch}{2.4}\tiny\CS}
\newcommand{\normalspacing}{\singlespacing}
\makeatother

\normalspacing



\newcommand{\up}{{\rm UP}}

\newcommand{\p}{{\rm P}}

\newcommand{\np}{{\rm NP}}

\newenvironment{proofs}{\noindent{\bf Proof.}\hspace*{1em}}{\qed\bigskip}

\newcommand{\sproofof}[1]{\noindent{\bf Proof of {#1}}\quad}
\newcommand{\eproofof}[1]{\noindent{\hspace*{0.1in} \hfil \hfill \mbox{\qed{} {#1}}}\quad}

\newcommand{\pair}[1]{\mathopen\langle{#1}\mathclose\rangle}

\newcommand{\sigmastar}{\ensuremath{\Sigma^\ast}}
\newcommand{\sigmaplus}{\ensuremath{\Sigma^+}}

\newcommand{\pisnotnp}{\ensuremath{\p\neq\np}}

\def\land{{\; \wedge \;}}
\def\lor{{\; \vee \;}}

\newcommand{\fp}{{\rm FP}}

\newcommand{\image}{\mbox{\rm{}image}}
\newcommand{\domain}{\mbox{\rm{}domain}}

\newcommand{\strong}{\mbox{$s$}}
\newcommand{\total}{\mbox{$t$}}
\newcommand{\commutative}{\mbox{$c$}}
\newcommand{\associative}{\mbox{$a$}}
\newcommand{\yes}{\mbox{\rm{}Y}}
\newcommand{\boldyes}{\mbox{\bf{}Y}}
\newcommand{\no}{\mbox{\rm{}N}}
\newcommand{\boldno}{\mbox{\bf{}N}}
\newcommand{\yesno}{\mbox{\rm{}$\ast$}}
\newcommand{\OWF}{\mbox{\rm{}OWF}}

\title{
Enforcing and Defying Associativity, Commutativity,
Totality, and Strong Noninvertibility for  One-Way Functions in
Complexity Theory\thanks{A preliminary version was presented 
at the 2005 ICTCS conference~\cite{hem-rot-sax:c:enforcing-defying-stac}.}
}

\author{
Lane A. Hemaspaandra\thanks{
URL: ${\tt www.cs.rochester.edu/u/lane}$.
Supported in part by 
NSF grant CCF-0426761, 
an Alexander von Humboldt Foundation 
TransCoop grant, and 
a Friedrich Wilhelm Bessel Research Award.
Work done in part while visiting 
Julius-Maximilians-Universit\"at W\"urzburg
and Heinrich-Heine-Universit\"at D\"usseldorf.} \\
Department of Computer Science \\
University of Rochester \\
Rochester, NY 14627, USA
\and 
J\"{o}rg Rothe\thanks{
URL: ${\tt ccc.cs.uni\mbox{-}duesseldorf.de/\tilde{~}rothe}$.
Supported in part by 
an Alexander von Humboldt Foundation 
TransCoop grant and 
DFG grants RO~1202/9-1, RO~1202/9-3 and RO~1202/11-1.
Work done in part while visiting
the University of Rochester
and 
Julius-Maximilians-Universit\"at W\"urzburg.} \\
Institut f\"ur Informatik \\
Heinrich-Heine-Universit\"at D\"usseldorf \\
40225 D\"usseldorf, Germany
\and 
Amitabh Saxena\thanks{
URL: ${\tt homepage.cs.latrobe.edu.au/asaxena}$.
}
\\Department of Computer Science and Computer Engineering \\
La Trobe University \\
Bundoora, VIC~3086, Australia
}

\date{
December 12, 2004; 
revised April 21, 2005
and November 1, 2007
}

\begin{document}

\sloppy

\maketitle

\begin{abstract}
\noindent
Rabi and Sherman~\cite{rab-she:j:aowf,rab-she:t-no-URL:aowf}
proved that the hardness of factoring is a sufficient condition
for there to exist one-way 
functions (i.e., p-time computable, honest, p-time noninvertible functions;
this paper is in the worst-case model, not the average-case model) 
that are total, commutative, and associative
but not strongly noninvertible.  In this paper we improve the 
sufficient condition to $\p \neq \np$.  

More generally,
in this paper we completely characterize which types of
one-way functions stand or fall together with (plain) one-way
functions---equivalently, stand or fall together with $\p \neq \np$.
We look at the four attributes used in Rabi and Sherman's seminal work
on algebraic properties of one-way functions (see
\cite{rab-she:j:aowf,rab-she:t-no-URL:aowf}) and subsequent
papers---strongness (of noninvertibility), totality, commutativity,
and associativity---and for each attribute, we allow it to be required
to hold, required to fail, or ``don't care.''  In this categorization
there are $3^4 = 81$ potential types of one-way functions.  We prove
that each of these $81$ feature-laden types stand or fall together with
the existence of (plain) one-way functions.  

\medskip


\end{abstract}

\section{Introduction}

\subsection{Motivation}

In this paper, we study properties of one-way functions, i.e.,
properties of functions that are easy to compute, but hard to invert.
One-way functions are important cryptographic primitives and are the
key building blocks in many cryptographic protocols.  Various models
to capture ``noninvertibility'' and, depending on the model used,
various candidates for one-way functions have been proposed.  The
notion of noninvertibility is usually based on the average-case
(where the ``average-case'' refers to the difficulty
of inversion)
complexity model 
(see, e.g., the 
book~\cite{gol:b:foundations}
and the references therein)
in cryptographic applications,
whereas
noninvertibility for complexity-theoretic one-way functions is usually
defined in the worst-case model (see the definitions 
of this paper).  Though the average-case model is
very important, we note that even the challenge of showing that any
type of one-way function exists in the ``less challenging''
worst-case model remains an open issue after many years of research.
It is thus natural to wonder, as a first step, what assumptions are
needed to create various types of complexity-theoretic
one-way functions.  In this paper, we seek to characterize 
this existence issue in terms
of class separations.  (In addition, we mention that 
the seminal work on associativity, commutativity,
and strong noninvertibility of 
one-way functions, which was done by
Rabi and Sherman~\cite{rab-she:t-no-URL:aowf,rab-she:j:aowf}
who also proposed concrete protocols to be based on such
one-way functions,
is itself in the worst-case model.)

Complexity-theoretic one-way functions of various sorts, and 
related notions, were
studied early on by, for example,
Berman~\cite{ber:thesis:iso}, 
Brassard, Fortune, and 
Hopcroft~\cite{bra-for-hop:t:one-way,bra:j:cryptocomplexity},
Ko~\cite{ko:j:operators}, and 
especially Grollmann and
Selman~\cite{gro-sel:j:complexity-measures}, 
and have been much
investigated ever since;
see, 
e.g.,~\cite{all-rub:j:print,wat:j:hardness-one-way,wat:b:one-way-survey,har-hem:j:one-way-iso,sel:j:one-way,rab-she:t-no-URL:aowf,gra:j:one-way,hem-rot-wec:c:easy-one-way-permutations,rab-she:j:aowf,hem-rot:j:aowf,bey-hem-hom-rot:j:aowf-survey,hem-rot:j:one-way,rot-hem:j:one-way,fen-for-nai-rog:j:inverse,hom-tha:j:one-way-permutations,hom:j:low-ambiguity-aowf,hem-pas-rot:j:strong-noninvertibility}.
The four properties of one-way functions to be investigated in this
paper are strongness, totality, commutativity, and associativity.
Intuitively, strong noninvertibility---a notion proposed by Rabi and
Sherman~\cite{rab-she:j:aowf,rab-she:t-no-URL:aowf} 
and more recently studied 
in~\cite{hem-rot:j:aowf,hom:j:low-ambiguity-aowf,hem-pas-rot:j:strong-noninvertibility}---means
that for a two-ary function, given some function value and one of the
corresponding arguments, it is hard to determine the other argument.
It has been known for decades that one-way functions exist if and only
if $\p \neq \np$.  But the Rabi-Sherman paper brought out the natural
issue of trying to understand what complexity-theoretic assumptions
characterized the existence of one-way functions with certain algebraic
properties.  Eventually,
Hemaspaandra and Rothe~\cite{hem-rot:j:aowf} proved that strong,
total, commutative, associative one-way functions exist if and only if
$\p \neq \np$.  (As mentioned earlier, 
one-way functions with these properties are the key
building blocks in Rabi, Rivest, and Sherman's cryptographic protocols
for secret-key agreement and for digital
signatures (see~\cite{rab-she:j:aowf,rab-she:t-no-URL:aowf}).)
The surprising work of Homan~\cite{hom:j:low-ambiguity-aowf}
both strengthens the results of Rabi and Sherman 
on the ambiguity that must be present in total, associative 
functions and proves that if 
one-to-one one-way functions exist, then there exist
strong, total, associative one-way functions having relatively
low ambiguity.

This paper provides a detailed study of the four properties of one-way
functions mentioned above.  For each possible combination of
possessing, not possessing, and being oblivious to possession of the
property, we study the question of whether such one-way functions can
exist.  Why should one be interested in knowing if a one-way function
possesses ``negative'' properties, such as noncommutativity?  
On one hand, negative 
properties can also have useful applications.  
For example, Saxena,
Soh, and 
Zantidis~\cite{sax-soh:c:signature-chaining-by-noncommutative-aowfs,sax-soh-zan:c:digital-cash}
propose authentication protocols for mobile agents and digital cash
with signature chaining that use as their key building blocks strong,
associative one-way functions for which commutativity in fact is a
disadvantage---though they need commutativity to not merely fail
but to fail far more often than is achieved in the failure constructions
of the present paper.
More generally and more importantly, given that 
complexity-theoretic one-way functions have 
already been studied for 
decades (see the citations above, going as far back 
as the 1970s),
it seems natural to try to 
understand and catalog
which types of one-way functions are created by, for example, simply
assuming $\p \neq \np$.   This paper does that completely
with respect to strongness, totality,
commutativity, and associativity.

\subsection{Summary of Our Results}

This paper is organized as follows.  In
Sections~\ref{sec:preliminaries} and~\ref{sec:groundwork}, we formally
define the notions and notation used, and we provide some basic lemmas
that allow us to drastically reduce the number of cases we have to
consider.  
We will state the full definitions later, but stated
merely intuitively,
a function is said to be strongly noninvertible if 
given the output and one argument one cannot efficiently
find a corresponding other argument;  and a function is said to 
be strong if it is polynomial-time computable, strongly 
noninvertible, and satisfies the natural honesty condition
related to strong noninvertibility (so-called s-honesty).
In Section~\ref{sec:strong-dontcare-if-strong-cases}, we
prove that the condition $\p \neq \np$ characterizes all $27$ cases
induced by one-way functions that are strong.
As a corollary, we also obtain a
$\p \neq \np$ characterization of all $27$ cases where one requires
one-way-ness but is oblivious to whether or not the functions are
strong.  In
Section~\ref{sec:nonstrong-cases}, we consider functions that are
required to be one-way but to not be strong.  We show that $\p
\neq \np$ characterizes all of these $27$ cases.  Thus,
$\p \neq \np$ characterizes all $81$ cases overall.  

\begin{table}[!tpb]
\centering
\begin{tabular}{|c|l|}
\hline
Properties              & 
                      $\p\neq\np$ characterization for
                      
 \\
$(\strong, \total, \commutative,\associative)$
                        &    this case is established by              \\ \hline
$(\no,\no,\no,\no)$     & 
Lemma~\ref{lem:nynn} +  Lemma~\ref{lem:eliminating-totality}   \\
$(\no,\no,\no,\yes)$    & 
Lemma~\ref{lem:foo-new} +  Lemma~\ref{lem:eliminating-totality}    \\
$(\no,\no,\yes,\no)$    & 
Lemma~\ref{lem:nyyn} +  Lemma~\ref{lem:eliminating-totality}   \\
$(\no,\no,\yes,\yes)$   & 
Lemma~\ref{lem:foo-new} +  Lemma~\ref{lem:eliminating-totality}   
                                     \\ \hline
$(\no,\yes,\no,\no)$    & 
\cite{hem-pas-rot:j:strong-noninvertibility}; see Lemma~\ref{lem:nynn} \\
$(\no,\yes,\no,\yes)$   & 
Lemma~\ref{lem:foo-new}
\\
$(\no,\yes,\yes,\no)$   & 
Lemma~\ref{lem:nyyn}                                           \\
$(\no,\yes,\yes,\yes)$  & 
Lemma~\ref{lem:foo-new}
\\ \hline
$(\yes,\no,\no,\no)$    & 
Lemma~\ref{lem:yynn} +  Lemma~\ref{lem:eliminating-totality}   \\
$(\yes,\no,\no,\yes)$   & 
Lemma~\ref{lem:yyny} +  Lemma~\ref{lem:eliminating-totality}   \\
$(\yes,\no,\yes,\no)$   & 
Lemma~\ref{lem:yyyn} +  Lemma~\ref{lem:eliminating-totality}   \\
$(\yes,\no,\yes,\yes)$  & 
Lemma~\ref{lem:yyyy} +  Lemma~\ref{lem:eliminating-totality}   \\ \hline
$(\yes,\yes,\no,\no)$   & 
Lemma~\ref{lem:yynn}                                           \\
$(\yes,\yes,\no,\yes)$  & 
Lemma~\ref{lem:yyny}                                           \\
$(\yes,\yes,\yes,\no)$  & 
Lemma~\ref{lem:yyyn}                                           \\
$(\yes,\yes,\yes,\yes)$ & 
\cite{hem-rot:j:aowf}, here restated as Lemma~\ref{lem:yyyy}           \\ \hline
\end{tabular}
\caption{Summary of 
support for the 16 key cases
\label{tab:results}
}
\end{table}

Table~\ref{tab:results} summarizes the support for 
our results for the $16$ key
cases in which each of the four properties considered is either
enforced or defied.\footnote{In light of the forthcoming
Lemma~\ref{lem:eliminating-yesno}, those cases in which one is
oblivious to whether some property holds follow immediately
from the cases stated in Table~\ref{tab:results}.}
Definition~\ref{def:scheme} provides the classification scheme used in
this table.  The left column of Table~\ref{tab:results} has $16$
quadruples of the form $(\strong, \total, \commutative,
\associative)$, where $\strong$ regards ``strong,'', $\total$ means
``total,'' $\commutative$ means ``commutative,'' and $\associative$
means ``associative.'' The variables $\strong$, $\total$,
$\commutative$, and $\associative$ take on a value from $\{\yes,
\no\}$, where $\yes$ means presence (i.e., ``yes''), and $\no$ means
absence (i.e., ``no'') of the given property.  The center column of
Table~\ref{tab:results} states the conditions characterizing the
existence of $(\strong, \total, \commutative, \associative)$-{\OWF}s,
and the right column of Table~\ref{tab:results} gives the references
to the proofs of the results stated.

\subsection{General Proof Strategy}

We do not attempt to brute-force all $81$ cases.  Rather, we seek to
turn the cases' structure and connectedness against themselves.  So,
in Section~\ref{sec:groundwork} we will reduce the~$81$ cases to
their~$16$ key cases that do not contain ``don't care'' conditions.
Then, also in Section~\ref{sec:groundwork}, we will show how to derive
the nontotal cases from the total cases, thus further reducing our
problem to~$8$ key cases.

As Corollary~\ref{cor:oblivious-to-strong-cases} and, especially,
much of Section~\ref{sec:nonstrong-cases} will show, even among the
$8$ key cases we share attacks, and find and exploit implications.

Thus, the proof in general consists both of specific
constructions---concrete 
realizations forcing 
given patterns of properties---and the framework that minimizes the
number of such constructions needed.

\section{Preliminaries and Notations}
\label{sec:preliminaries}

Fix the alphabet $\Sigma = \{0,1\}$.  The set of strings over $\Sigma$
is denoted by~$\sigmastar$.  Let $\varepsilon$ denote the empty
string.  Let $\sigmaplus = \sigmastar - \{\varepsilon\}$.  For any
string $x \in \sigmastar$, let $|x|$ denote the length of~$x$.  Let
$\pair{\cdot, \cdot} : \sigmastar \times \sigmastar \rightarrow\,
\sigmastar$ be some standard pairing function, that is, some total,
polynomial-time computable bijection that has polynomial-time
computable inverses and is nondecreasing in each argument when the
other argument is fixed.  
Let $\fp$ denote the class of polynomial-time
computable functions (this includes both total and nontotal functions).
This paper focuses completely on mappings from $\sigmastar \times
\sigmastar$ to $\sigmastar$ (they are allowed to be many-to-one and 
they are allowed to be nontotal, i.e., they may map many distinct pairs
of strings from $\sigmastar \times \sigmastar$ to one and the same
string in~$\sigmastar$, and they need not be defined for all pairs in
$\sigmastar \times \sigmastar$).  (The study of $2$-argument one-way 
functions of course is needed if associativity and commutativity are
to be studied.)  For each function~$f$, let $\domain(f)$ denote the set
of input pairs on which $f$ is defined, and denote the image of $f$ by
$\image(f)$.

Definition~\ref{def:one-way-function} presents the standard notion of a
(complexity-theoretic, many-one) one-way function, 
suitably
tailored to the case of two-ary functions in the standard way; 
see~\cite{rab-she:t-no-URL:aowf,rab-she:j:aowf,hem-rot:j:aowf,hom:j:low-ambiguity-aowf,hem-pas-rot:j:strong-noninvertibility}.
(For general introductions to or surveys on one-way functions,
see~\cite{sel:j:one-way},
\cite{bey-hem-hom-rot:j:aowf-survey},
and~\cite[Chapter~2]{hem-ogi:b:companion}.
For general background on complexity see, e.g.,
\cite{hem-ogi:b:companion,bov-cre:b:complexity}.)
Our one-way functions are based on noninvertibility in the
worst-case model, as opposed to noninvertibility in the average-case
model that is more appealing for cryptographic applications.  The
notion of honesty in Definition~\ref{def:one-way-function} below is
needed in order to preclude functions from being noninvertible simply due to
the trivial reason that some family of images lacks 
polynomially short preimages.

\begin{definition}[One-Way Function]\quad
\label{def:one-way-function}
Let $\sigma$ be a function (it may be either total or nontotal) mapping from
  $\sigmastar \times \sigmastar$ to $\sigmastar$.
\begin{enumerate}
\item We say $\sigma$ is {\em honest\/} if and only if there exists a
  polynomial $p$ such that for each $z \in \image(\sigma)$, there
  exists a pair $(x,y) \in \domain(\sigma)$ such that $\sigma(x,y) =
  z$ and $|x| + |y| \leq p(|z|)$.

\item We say $\sigma$ is {\em (polynomial-time) noninvertible\/} if
  and only if there exists no function $f$ in $\fp$ such that for all
  $z \in \image(\sigma)$, we have $\sigma(f(z)) = z$.

\item We say $\sigma$ is a {\em one-way function\/} if and only if
  $\sigma$ is polynomial-time computable, honest, and noninvertible.
\end{enumerate}
\end{definition}

The four properties of one-way functions that we will study in this
paper are strongness, totality, commutativity, and associativity.  A
function $\sigma$ mapping from $\sigmastar \times \sigmastar$ to
$\sigmastar$ is said to be {\em total\/} if and only if $\sigma$ is
defined for each pair in $\sigmastar \times \sigmastar$, and is said 
to be {\em nontotal\/} if it is not total.  We say that
a function is {\em partial\/} if it is either total or nontotal; this
says nothing, but makes it clear that we are not demanding that the 
function be total.

We now define the remaining three properties.  Rabi, Rivest, and
Sherman (see~\cite{rab-she:j:aowf,rab-she:t-no-URL:aowf})
introduced the notion of strongly noninvertible associative
one-way functions (strong AOWFs, for short).  Rivest and Sherman
(as attributed in~\cite{rab-she:j:aowf,rab-she:t-no-URL:aowf}) 
designed cryptographic
protocols for two-party
secret-key agreement and Rabi and Sherman designed cryptographic
protocols for digital signatures, both of which need strong, total
AOWFs as their key building blocks.  They also sketch protocols for
multiparty secret-key agreement that required strong, total, commutative
AOWFs.  Strong (and sometimes total and
commutative) AOWFs have been intensely studied
in~\cite{hem-rot:j:aowf,bey-hem-hom-rot:j:aowf-survey,hom:j:low-ambiguity-aowf,hem-pas-rot:j:strong-noninvertibility}.

Though Rabi and Sherman's~\cite{rab-she:j:aowf} notion of associativity
is meaningful for total functions, it is not meaningful for
nontotal two-ary functions, as has been noted and discussed
in~\cite{hem-rot:j:aowf}.  Thus, we here follow Hemaspaandra and
Rothe's~\cite{hem-rot:j:aowf} notion of associativity, which is
appropriate for both total and nontotal two-ary functions, and is designed
as an analog to Kleene's 1952~\cite{kle:b:metamathematics} notion
of complete equality of partial functions.

\begin{definition}[Associativity and Commutativity]\quad
\label{def:associative-commutative}
Let $\sigma$ be any partial function mapping from
  $\sigmastar \times \sigmastar$ to $\sigmastar$.  Extend $\sigmastar$
  by $\Gamma = \sigmastar \cup \{\bot\}$, where $\bot$ is a special
  symbol indicating, in the usage ``$\sigma(x,y) = \bot$,'' that
  $\sigma$ is not defined for the pair $(x,y)$.  Define an extension
  $\widehat{\sigma}$ of $\sigma$, which maps from $\Gamma \times
  \Gamma$ to $\Gamma$, as follows:
\begin{equation}
  \widehat{\sigma}(x,y) = \left\{ 
\begin{array}{ll} 
\sigma(x,y)  &
  \mbox{if $x \neq \bot$ and $y \neq \bot$ and $(x,y) \in \domain(\sigma)$} \\ 
  \bot & \mbox{otherwise.}
\end{array} 
\right.
\end{equation}
\begin{enumerate}
\item We say $\sigma$ is {\em associative\/} if and only if for each
$x,y,z \in \sigmastar$,
$\widehat{\sigma}(\widehat{\sigma}(x,y),z)
 = \widehat{\sigma}(x,\widehat{\sigma}(y,z))$.

\item We say $\sigma$ is {\em commutative\/} if and only if for each
$x,y \in \sigmastar$,
$\widehat{\sigma}(x,y) = \widehat{\sigma}(y,x)$.
\end{enumerate}
\end{definition}

Informally speaking, strong noninvertibility 
(see~\cite{rab-she:j:aowf,rab-she:t-no-URL:aowf}) means that even if a
function value and one of the corresponding two arguments are given,
it is hard to compute the other argument.  It is known that, unless
$\p = \np$, some noninvertible functions are not strongly 
noninvertible~\cite{hem-pas-rot:j:strong-noninvertibility}.
And, perhaps counterintuitively, it is known that, unless
$\p = \np$, some strongly noninvertible functions are not
noninvertible~\cite{hem-pas-rot:j:strong-noninvertibility}.  That is,
unless $\p = \np$, strong noninvertibility does not imply
noninvertibility.  Strong noninvertibility requires a
variation of honesty that is dubbed s-honesty
in~\cite{hem-pas-rot:j:strong-noninvertibility}.
The notion defined now, as ``strong (function)'' in
Definition~\ref{def:strong-one-way-function},
is in the literature
typically called a ``strong one-way function.''  This is quite natural.
However, 
to avoid any possibility of 
confusion as to when we refer to that and 
when we refer to the notion of a ``one-way function'' 
(see Definition~\ref{def:one-way-function};  as will be mentioned 
later, neither of these notions necessarily implies the other), 
we will throughout
this paper simply call the notion below ``strong'' 
or ``a strong function,'' rather than ``strong one-way function.''

\begin{definition}[Strong Function]\quad
\label{def:strong-one-way-function}
Let $\sigma$ be any partial function mapping from
  $\sigmastar \times \sigmastar$ to $\sigmastar$.
\begin{enumerate}
\item We say $\sigma$ is {\em s-honest\/} if and only if there exists a
  polynomial $p$ such that the following two conditions are true:
\begin{enumerate}
\item[(a)] For each $x,z \in \sigmastar$ with $\sigma(x,y) = z$ for
some $y \in \sigmastar$, there exists some string $\hat{y} \in
\sigmastar$ such that
$\sigma(x,\hat{y}) = z \mbox{ and } |\hat{y}| \leq p(|x| + |z|)$.

\item[(b)] For each $y,z \in \sigmastar$ with $\sigma(x,y) = z$ for
some $x \in \sigmastar$, there exists some string $\hat{x} \in
\sigmastar$ such that
$\sigma(\hat{x},y) = z \mbox{ and } |\hat{x}| \leq p(|y| + |z|)$.
\end{enumerate}

\item We say $\sigma$ is {\em (polynomial-time) invertible with
    respect to the first argument\/} if and only if there exists an
    inverter $g_1 \in \fp$ such that for every string $z \in
    \image(\sigma)$ and for all $x,y \in \sigmastar$ with $(x,y) \in
    \domain(\sigma)$ and $\sigma(x,y) = z$,
\[
\sigma(x, g_1(\pair{x,z})) = z.
\]

\item We say $\sigma$ is {\em (polynomial-time) invertible with
    respect to the second argument\/} if and only if there exists an
    inverter $g_2 \in \fp$ such that for every string $z \in
    \image(\sigma)$ and for all $x,y \in \sigmastar$ with $(x,y) \in
    \domain(\sigma)$ and $\sigma(x,y) = z$,
\[
\sigma(g_2(\pair{y,z}), y) = z.
\]

\item We say $\sigma$ is {\em strongly noninvertible\/} if and only if
    $\sigma$ is neither invertible with respect to the first argument
    nor invertible with respect to the second argument.

\item We say $\sigma$ is {\em strong\/} if and only
  if $\sigma$ is polynomial-time computable, s-honest, and strongly
  noninvertible.
\end{enumerate}
\end{definition}

In this paper, we will look at the $3^4 = 81$ categories of one-way
functions that one can get by requiring the properties 
strong/total/commutative/associative to either: hold, fail, or ``don't care.''
For each, we will try to characterize whether such one-way functions exist.

We now define a classification scheme suitable to capture all possible
combinations of these four properties of one-way functions.

\begin{definition}[Classification Scheme for One-Way Functions]\quad
\label{def:scheme}
For each $\strong, \total, \commutative, \associative \in
\{\yes,\allowbreak \no, \yesno\}$, we say that a partial function 
$\sigma : \sigmastar \times \sigmastar \rightarrow \sigmastar$
is an {\em $(\strong,\total, \commutative, \associative)$ one-way function 
(an $(\strong, \total, \commutative, \associative)$-{\OWF}, for short)\/} 
if and only if
all the following hold:
$\sigma$ is a one-way function,
if $\strong = \yes$ then $\sigma$ is strong,
if $\strong = \no$ then $\sigma$ is not strong,
 if $\total = \yes$ then $\sigma$ is a total function,
 if $\total = \no$ then $\sigma$ is a nontotal function,
 if $\commutative = \yes$ then $\sigma$ is a commutative function,
 if $\commutative = \no$ then $\sigma$ is a noncommutative function,
if $\associative = \yes$ then $\sigma$ is an associative function, and
 if $\associative = \no$ then $\sigma$ is a nonassociative function.
\end{definition}

For example, a function is a $(\yes, \yes, \yes, \yes)$-{\OWF} exactly if
it is a strong, total, commutative, associative one-way function.  
And note that, under this definition,
whenever a setting is $\yesno$, we don't place any 
restriction as to 
whether the corresponding
property holds or fails to hold---that is, $\yesno$ is a 
``don't care'' designator.
For example, a
function is a $(\yesno, \yes, \yesno, \yesno)$-{\OWF} exactly if
it is a total one-way function.  Of course, all 
$(\yes, \yes, \yes, \yes)$-{\OWF}s are 
$(\yesno, \yes, \yesno, \yesno)$-{\OWF}s.  That is, our $81$ classes
do not seek to partition, but rather to allow all possible simultaneous
settings and ``don't care''s for these four properties.  However, the $16$
such classes with no stars are certainly pairwise disjoint.

\section{Groundwork: Reducing the Cases}
\label{sec:groundwork}

In this section, we show how to tackle our ultimate goal, stated as
Goal~\ref{goal:main-result} below, by drastically reducing the number
of cases that are relevant among the $81$ possible cases.

\begin{goal}
\label{goal:main-result}
For each $\strong, \total, \commutative, \associative \in \{\yes, \no,
\yesno\}$, characterize the existence of $(\strong, \total,
\commutative, \associative)$-{\OWF}s in terms of some suitable
complexity-theoretic condition.
\end{goal}

Since $\yesno$ is a ``don't care,'' for a given $\yesno$ position the
characterization that holds with that $\yesno$ is simply the ``or'' of
the characterizations that hold with each of $\yes$ and $\no$
substituted for the~$\yesno$.  For example, clearly there exist
$(\yes, \yes, \yes, \yesno)$-{\OWF}s if and only if either there exist
$(\yes, \yes, \yes, \yes)$-{\OWF}s or there exist $(\yes, \yes, \yes,
\no)$-{\OWF}s.  And cases with more than one $\yesno$ can be
``unwound'' by repeating this.  So, to characterize all $81$ cases,
it suffices to characterize the $16$ cases stated in
Table~\ref{tab:results}.

\begin{lemma}
\label{lem:eliminating-yesno}
\begin{enumerate}
\item For each $\total, \commutative, \associative \in \{\yes, \no,
\yesno\}$, there exist
$(\yesno,\total,\commutative,\associative)$-{\OWF}s if and only if
either there exist $(\yes,\total,\commutative,\associative)$-{\OWF}s
or there exist $(\no,\total,\commutative,\associative)$-{\OWF}s.

\item For each $\strong, \commutative, \associative \in \{\yes, \no,
\yesno\}$, there exist
$(\strong,\yesno,\commutative,\associative)$-{\OWF}s if and only if
either there exist $(\strong,\yes,\commutative,\associative)$-{\OWF}s
or there exist $(\strong,\no,\commutative,\associative)$-{\OWF}s.

\item For each $\strong, \total, \associative \in \{\yes, \no,
\yesno\}$, there exist $(\strong,\total,\yesno,\associative)$-{\OWF}s
if and only if either there exist
$(\strong,\total,\yes,\associative)$-{\OWF}s or there exist
$(\strong,\total,\no,\associative)$-{\OWF}s.

\item For each $\strong, \total, \commutative \in \{\yes, \no,
\yesno\}$, there exist $(\strong,\total,\commutative,\yesno)$-{\OWF}s
if and only if either there exist
$(\strong,\total,\commutative,\yes)$-{\OWF}s or there exist
$(\strong,\total,\commutative,\no)$-{\OWF}s.
\end{enumerate}
\end{lemma}

It is well known
(see~\cite{bal-dia-gab:b:sctI:95} and Proposition~1
of~\cite{sel:j:one-way}) that $\p \neq \np$ if and only if
$(\yesno,\yesno,\yesno,\yesno)$-{\OWF}s exist, i.e., $\p \neq \np$ if
and only if there exist one-way functions, regardless of whether or
not they possess any of the four properties.  So, in the upcoming
proofs, we will often focus on just showing that $\p \neq \np$ implies
the given type of {\OWF} exists.

\begin{lemma}
\label{lem:eliminating-one-direction}
For each $\strong, \total, \commutative, \associative \in \{\yes,
\no, \yesno\}$, if there are $(\strong, \total, \commutative,
\associative)$-{\OWF}s then $\p \neq \np$.
\end{lemma}

Next, we show that all cases involving nontotal one-way functions can
be easily reduced to the corresponding cases involving total one-way
functions.  Thus, we have eliminated the eight ``nontotal'' of the
remaining $16$ cases, provided we can solve the eight ``total'' cases.

\begin{lemma}
\label{lem:eliminating-totality}
For each $\strong, \commutative, \associative \in \{\yes, \no\}$, if
there exists an $(\strong, \yes, \commutative, \associative)$-{\OWF},
then there exists an $(\strong, \no, \commutative,
\associative)$-{\OWF}.
\end{lemma}

\begin{proofs}
Fix any $\strong, \commutative, \associative \in \{\yes, \no\}$, and
let $\sigma$ be any given $(\strong, \yes, \commutative,
\associative)$-{\OWF}.  For each string $w \in \sigmastar$, let $w^+$
denote the successor of $w$ in the standard lexicographic ordering
of~$\sigmastar$, and for each string $w \in \sigmaplus$, let $w^-$
denote the predecessor of $w$ in the standard lexicographic ordering
of~$\sigmastar$.

Define a function $\rho : \sigmastar \times \sigmastar
\rightarrow \sigmastar$ by
\[
\rho(x,y) = \left\{
\begin{array}{ll}
(\sigma(x^-,y^-))^+ & \mbox{ if $x \neq \varepsilon \neq y$} \\
\mbox{undefined} & \mbox{ otherwise.}
\end{array}
\right.
\]
Note that $\rho$ is nontotal, since it is not defined on the pair
$(\varepsilon, \varepsilon)$.  It is a matter of routine to check that
$\rho$ is a one-way function, i.e., polynomial-time computable,
honest, and noninvertible.  It remains to show that $\rho$ inherits
all the other properties from $\sigma$ as well.  To this end, we show
the following claim.

\begin{claim}
\label{cla:eliminating-totality}
\begin{enumerate}
\item $\sigma$ is commutative if and only if $\rho$ is commutative.
\item $\sigma$ is associative if and only if $\rho$ is associative.
\item $\sigma$ is strong if and only if $\rho$ is 
strong.
\end{enumerate}
\end{claim}

\sproofof{Claim~\ref{cla:eliminating-totality}} We check these
properties separately.
\begin{enumerate}
\item {\bf Commutativity:} Suppose that $\sigma$ is
commutative.  Given any strings $x, y \in \sigmastar$, if $x =
\varepsilon$ or $y = \varepsilon$, then both $\rho(x,y)$ and
$\rho(y,x)$ are undefined.  If $x \neq \varepsilon \neq y$, then the
commutativity of $\sigma$ implies that
\[
\rho(x,y) = (\sigma(x^-,y^-))^+ = (\sigma(y^-,x^-))^+ = \rho(y,x).
\]
So $\widehat{\rho}(x,y) = \widehat{\rho}(y,x)$.  By
Definition~\ref{def:associative-commutative}, $\rho$ is commutative.

Conversely, suppose that $\sigma$ is noncommutative.  Since $\sigma$
is total, we don't have to worry about holes in the domain
of~$\sigma$.  Let $a$ and $b$ be fixed strings in $\sigmastar$ such
that $\sigma(a,b) \neq \sigma(b,a)$.  It follows that
\[
\rho(a^+, b^+) \neq \rho(b^+, a^+) .
\]
Thus, $\rho$ is noncommutative.

\item {\bf Associativity:} Suppose that $\sigma$ is
associative.  Let $x$, $y$, and $z$ be any strings in~$\sigmastar$.  If $x =
\varepsilon$ or $y = \varepsilon$ or $z = \varepsilon$, then both
$\rho(x,\rho(y,z))$ and $\rho(\rho(x,y),z)$ are undefined.  If none of
$x$, $y$, and $z$ equals the empty string, then the associativity of
$\sigma$ implies
\begin{eqnarray*}
\rho(x,\rho(y,z)) & = & (\sigma(x^-,\sigma(y^-,z^-)))^+ \\ 
                  & = & (\sigma(\sigma(x^-,y^-),z^-))^+ \\
                  & = & \rho(\rho(x,y),z).
\end{eqnarray*}
So $\widehat{\rho}(x,\widehat{\rho}(y,z)) =
\widehat{\rho}(\widehat{\rho}(x,y),z)$.  By
Definition~\ref{def:associative-commutative}, $\rho$ is associative.

Conversely, suppose that $\sigma$ is nonassociative.  Let $a$, $b$,
and $c$ be fixed strings in $\sigmastar$ such that
$\sigma(a,\sigma(b,c)) \neq \sigma(\sigma(a,b),c)$.  Since $\sigma$ is
total, each of $\sigma(a,b)$, $\sigma(b,c)$, $\sigma(a,\sigma(b,c))$,
and $\sigma(\sigma(a,b),c)$ is defined.  So
\begin{eqnarray*}
(\rho(a^+,\rho(b^+,c^+)))^- & = & \sigma(a,\sigma(b,c)) \\ 
                         & \neq & \sigma(\sigma(a,b),c) \\
                            & = & (\rho(\rho(a^+,b^+),c^+))^- ,
\end{eqnarray*}
which implies $\rho(a^+,\rho(b^+,c^+)) \neq \rho(\rho(a^+,b^+),c^+)$.
Thus,
\[
\widehat{\rho}(a^+,\widehat{\rho}(b^+,c^+)) \neq
\widehat{\rho}(\widehat{\rho}(a^+,b^+),c^+).
\]
By Definition~\ref{def:associative-commutative}, $\rho$ is
nonassociative.

\item {\bf Strongness:} First, we note that $\sigma$ is
s-honest if and only if $\rho$ is s-honest.  Let $p$ be some
polynomial witnessing the s-honesty of $\sigma$ as per
Definition~\ref{def:strong-one-way-function}:
\begin{enumerate}
\item[(a)] For each $x,z \in \sigmastar$ with $\sigma(x,y) = z$ for
some $y \in \sigmastar$, there exists some string $\hat{y} \in
\sigmastar$ such that $\sigma(x,\hat{y}) = z \mbox{ and } |\hat{y}|
\leq p(|x| + |z|)$.

\item[(b)] For each $y,z \in \sigmastar$ with $\sigma(x,y) = z$ for
some $x \in \sigmastar$, there exists some string $\hat{x} \in
\sigmastar$ such that $\sigma(\hat{x},y) = z \mbox{ and } |\hat{x}|
\leq p(|y| + |z|)$.
\end{enumerate}
Since $\rho$ shifts the arguments and the function value of $\sigma$
just by one position in the lexicographic ordering on~$\sigmastar$,
the polynomial $q(n) = p(n)+1$ witnesses the s-honesty of~$\rho$.  The
converse is proven analogously.

Now, we show that $\sigma$ is strongly noninvertible if and only if
$\rho$ is strongly noninvertible.  Suppose that $\sigma$ is invertible
with respect to the first argument via some inverter~$g_1$ in~$\fp$.
That is, for each string $z \in \image(\sigma)$ and for all $x,y \in
\sigmastar$ with $(x,y) \in \domain(\sigma)$ and $\sigma(x,y) = z$, we
have
\[
\sigma(x, g_1(\pair{x,z})) = z.
\]
From $g_1$ we construct an inverter $f_1 \in \fp$ that inverts $\rho$
with respect to the first argument as follows.  Let $z$ be any string
in $\image(\rho)$, and let $x,y \in \sigmastar$ be any strings such
that $(x,y) \in \domain(\rho)$ and $\rho(x,y) = z$.  Given
$\pair{x,z}$, $f_1$ computes $(g_1(\pair{x^-,z^-}))^+$.  Note that
$\rho$ never maps to the empty string, so $z \neq \varepsilon$ and
$z^-$ is well-defined.  Similarly, $x \neq \varepsilon$ because $(x,y)
\in \domain(\rho)$, so $x^-$ is well-defined.  Thus,
\[
\rho(x,f_1(\pair{x,z})) =  \rho(x,(g_1(\pair{x^-,z^-}))^+) = z .
\]
Similarly, an inverter with respect to the second argument can be
built for $\rho$ given one for~$\sigma$.  

Conversely, given an inverter for $\rho$ with respect to the first
(respectively, second) argument, an inverter for $\sigma$ with respect
to the first (respectively, second) argument can be constructed by
reverting the shifting above.  Thus, if $\rho$ is not strongly
noninvertible, neither is~$\sigma$.
\end{enumerate}
\eproofof{Claim~\ref{cla:eliminating-totality}} 

This completes the proof of Lemma~\ref{lem:eliminating-totality}.
\end{proofs}

Lemmas~\ref{lem:eliminating-yesno},
\ref{lem:eliminating-one-direction},
and~\ref{lem:eliminating-totality} imply that it suffices to deal with
only the ``total'' cases.  That is, to achieve
Goal~\ref{goal:main-result}, it would be enough to show that if $\p
\neq \np$ then each of the following eight types of one-way functions
exist:
$(\yes,\yes,\yes,\yes)$-{\OWF}s, 
$(\yes,\yes,\yes,\no)$-{\OWF}s,  
$(\yes,\yes,\no,\yes)$-{\OWF}s,
$(\yes,\yes,\no,\no)$-{\OWF}s, 
$(\no,\yes,\yes,\yes)$-{\OWF}s, 
$(\no,\yes,\yes,\no)$-{\OWF}s, 
$(\no,\yes,\no,\yes)$-{\OWF}s, 
and
$(\no,\yes,\no,\no)$-{\OWF}s.   
In the following sections, we will study each of these cases.

\section{Strongness and Being Oblivious to Strongness: {\boldmath $(\boldyes,\total,\commutative,\associative)$}-{OWF}s and {\boldmath $(\yesno,\total,\commutative,\associative)$}-{OWF}s}
\label{sec:strong-dontcare-if-strong-cases}

In this section, we consider the ``strong''-is-required cases and those cases
where the property of strongness is a ``don't care'' issue. 
We start with the $27$
``strong'' cases.  Theorem~\ref{thm:strong-cases} below characterizes
each of these cases by the condition $\p \neq \np$.  The proof of
Theorem~\ref{thm:strong-cases} follows from the upcoming
Lemmas~\ref{lem:yyyy} through~\ref{lem:yynn}, via
Lemmas~\ref{lem:eliminating-yesno},
\ref{lem:eliminating-one-direction},
and~\ref{lem:eliminating-totality}.

\begin{theorem}
\label{thm:strong-cases}
For each $\total, \commutative, \associative \in \{\yes, \no,
\yesno\}$, there exist $(\yes, \total, \commutative,
\associative)$-{\OWF}s if and only if $\p \neq \np$.
\end{theorem}

Lemma~\ref{lem:yyyy} is already known from Hemaspaandra and Rothe's
work~\cite{hem-rot:j:aowf}.

\begin{lemma}
\label{lem:yyyy}
If $\p \neq \np$ then there exist $(\yes,\yes,\yes,\yes)$-{\OWF}s.
\end{lemma}

The equivalence (due to~\cite{hem-rot:j:aowf},
and following immediately from 
Lemma~\ref{lem:yyyy} in light 
of Lemma~\ref{lem:eliminating-one-direction})
of $\p \neq \np$ and the existence of
$(\yes,\yes,\yes,\yes)$-{\OWF}s will be exploited 
in the upcoming proofs of
Lemmas~\ref{lem:yyyn}, \ref{lem:yyny}, and~\ref{lem:yynn}. 
That is, in these
proofs, we start from a strong, total, commutative, associative
one-way function.

\begin{lemma}
\label{lem:yyyn}
If $\p \neq \np$ then there exist $(\yes,\yes,\yes,\no)$-{\OWF}s.
\end{lemma}

\begin{proofs}
By Lemmas~\ref{lem:eliminating-one-direction} and~\ref{lem:yyyy}, the
condition $\p \neq \np$ is equivalent to the existence of some
$(\yes,\yes,\yes,\yes)$-{\OWF}, call it~$\sigma$.  Recall from the
proof of Lemma~\ref{lem:eliminating-totality} that, in the standard
lexicographic ordering of~$\sigmastar$, $w^+$ denotes the successor of
$w \in \sigmastar$ and $w^-$ denotes the predecessor of $w \in
\sigmaplus$.  We use the following shorthand: For $w \in \sigmastar$,
let $w^{2+} = (w^{+})^+$, and for $w \in \sigmastar$ with $w \not\in
\{\varepsilon, 0\}$, let $w^{2-} = (w^{-})^-$.
Define a function $\rho : \sigmastar \times \sigmastar
\rightarrow \sigmastar$ by
\[
\rho(x,y) = \left\{
\begin{array}{ll}
\varepsilon & \mbox{ if $x = y = 0$} \\
0           & \mbox{ if $x = y = \varepsilon$} \\
\varepsilon & \mbox{ if $(x = \varepsilon \land y = 0) \lor
                         (x = 0 \land y = \varepsilon)$} \\
\varepsilon & \mbox{ if $\{x,y\} \cap \{\varepsilon,0\} \neq \emptyset \land 
   \{x,y\} \cap \left(\sigmastar - \{\varepsilon,0\}\right) \neq \emptyset$} \\
(\sigma(x^{2-},y^{2-}))^{2+} & \mbox{ otherwise.}
\end{array}
\right.
\]

It is easy to see that $\rho$ is one-way, strong, total, and
commutative.  This fact can be seen to follow from the construction of
$\rho$ and from $\sigma$ having all these properties. However, $\rho$
is not an associative function, since
$\rho(\varepsilon,\rho(\varepsilon,0)) = 0 \neq \varepsilon =
\rho(\rho(\varepsilon,\varepsilon),0)$.

Thus, $\rho$ is a $(\yes,\yes,\yes,\no)$-{\OWF}.
\end{proofs}

\begin{lemma}
\label{lem:yyny}
If $\p \neq \np$ then there exist $(\yes,\yes,\no,\yes)$-{\OWF}s.
\end{lemma}

\begin{proofs}
Assuming $\p \neq \np$.  By Lemma~\ref{lem:yyyy}, let $\sigma$ be a
$(\yes,\yes,\yes,\yes)$-{\OWF}.  Define a function $\rho :
\sigmastar \times \sigmastar \rightarrow \sigmastar$ by
\[
\rho(x,y) = \left\{
\begin{array}{ll}
y                            & \mbox{ if $x, y \in \{0,1\}$} \\
(\sigma(x^{3-},y^{3-}))^{3+} & \mbox{ if $x \not\in \{\varepsilon,0,1\} \land 
                                          y \not\in \{\varepsilon,0,1\})$} \\
\varepsilon                  & \mbox{ otherwise,}
\end{array}
\right.
\]
where we use the following shorthand: Recall from the
proof of Lemma~\ref{lem:eliminating-totality} that, in the standard
lexicographic ordering of~$\sigmastar$, $w^+$ denotes the successor of
$w \in \sigmastar$ and $w^-$ denotes the predecessor of $w \in
\sigmaplus$.  For $w \in \sigmastar$, let
$w^{3+} = ((w^{+})^{+})^+$, and for $w \in \sigmastar$ with $w \not\in
\{\varepsilon, 0, 1\}$, let $w^{3-} = ((w^{-})^{-})^-$.

It is easy to see, given the fact that $\sigma$ is a
$(\yes,\yes,\yes,\yes)$-{\OWF}, that $\rho$ is a strongly noninvertible,
s-honest,  total one-way function.  However,
unlike~$\sigma$, $\rho$ is noncommutative, since
\[
\rho(0,1) = 1 \neq 0 = \rho(1,0).
\]
To see that~$\rho$, just like~$\sigma$, is associative, let three
arbitrary strings be given, say~$a$, $b$, and~$c$.  Distinguish the
following cases:
\begin{description}
\item[{\bf Case~1:}] {\it Each of $a$, $b$, and~$c$ is
a member of $\{0,1\}$.} Then, associativity follows from the
definition of~$\rho$:
\[
\rho(a,\rho(b,c)) = \rho(a,c) = c = \rho(b,c) = \rho(\rho(a,b),c) .
\]

\item[{\bf Case~2:}] {\it None of $a$, $b$, and~$c$ is
a member of $\{\varepsilon,0,1\}$.}  Then the associativity of $\rho$
follows immediately from the associativity of~$\sigma$.  That is,
\begin{eqnarray*}
\rho(a,\rho(b,c)) & = & \rho(a,(\sigma(b^{3-},c^{3-}))^{3+}) \\ 
                  & = & (\sigma(a^{3-},\sigma(b^{3-},c^{3-})))^{3+} \\ 
                  & = & (\sigma(\sigma(a^{3-},b^{3-}),c^{3-}))^{3+} \\
                  & = & \rho((\sigma(a^{3-},b^{3-}))^{3+},c) \\
                  & = & \rho(\rho(a,b),c).
\end{eqnarray*}
Note here that both $(\sigma(a^{3-},b^{3-}))^{3+}$ and
$(\sigma(b^{3-},c^{3-}))^{3+}$ are strings that are not 
members of 
$\{\varepsilon,0,1\}$.

\item[{\bf Case~3:}] {\it At least one of $a$, $b$, and~$c$
is not a member of $\{0,1\}$, and at least one of $a$, $b$, and~$c$
is a member of $\{\varepsilon,0,1\}$.}  In this case, it follows from
the definition of~$\rho$ that
\[
\rho(a,\rho(b,c)) = \varepsilon = \rho(\rho(a,b),c) .
\]
\end{description}
Thus, $\rho$ is a $(\yes,\yes,\no,\yes)$-{\OWF}.
\end{proofs}

\begin{lemma}
\label{lem:yynn}
If $\p \neq \np$ then there are $(\yes,\yes,\no,\no)$-{\OWF}s.
\end{lemma}

\begin{proofs}
Assume $\p \neq \np$.  By Lemma~\ref{lem:yyyy}, let $\sigma$ be a
$(\yes,\yes,\yes,\yes)$-{\OWF}.  Define a function $\rho :
\sigmastar \times \sigmastar \rightarrow \sigmastar$ by
\[
\rho(x,y) = \left\{
\begin{array}{ll}
\varepsilon & \mbox{ if $x = y = 0$} \\
0           & \mbox{ if $x = y = \varepsilon$} \\
\varepsilon & \mbox{ if $x = \varepsilon \land y = 0$} \\
0           & \mbox{ if $x = 0 \land y = \varepsilon$} \\
\varepsilon & \mbox{ if $\{x,y\} \cap \{\varepsilon,0\} \neq \emptyset \land 
   \{x,y\} \cap \left(\sigmastar - \{\varepsilon,0\}\right) \neq \emptyset$} \\
(\sigma(x^{2-},y^{2-}))^{2+} & \mbox{ otherwise.}
\end{array}
\right.
\]
Again, it follows from the properties of $\sigma$ and the construction
of $\rho$ that $\rho$ is one-way, strong, and total.
However, $\rho$ is not commutative, since
\[
\rho(\varepsilon,0) = \varepsilon \neq 0 = \rho(0,\varepsilon) .
\]
Furthermore, 
$\rho$ is not associative, since
\[
\rho(\varepsilon,\rho(\varepsilon,0)) =  0 \neq \varepsilon = 
\rho(\rho(\varepsilon,\varepsilon),0) .
\]
Thus, $\rho$ is a $(\yes,\yes,\no,\no)$-{\OWF}.
\end{proofs}

Next, we note Corollary~\ref{cor:oblivious-to-strong-cases}, which
follows immediately from Theorem~\ref{thm:strong-cases} via
Lemmas~\ref{lem:eliminating-yesno}
and~\ref{lem:eliminating-one-direction}.  That is, in light of
Lemmas~\ref{lem:eliminating-yesno}
and~\ref{lem:eliminating-one-direction},
Theorem~\ref{thm:strong-cases} provides also a $\p \neq \np$
characterization of all $27$ cases where one requires one-way-ness
but is oblivious to whether or not the functions are guaranteed to be
strong.

\begin{corollary}
\label{cor:oblivious-to-strong-cases}
For each $\total, \commutative, \associative \in \{\yes, \no,
\yesno\}$, there are $(\yesno, \total, \commutative,
\associative)$-{\OWF}s if and only if $\p \neq \np$.
\end{corollary}

\section{Nonstrongness: {\boldmath $(\boldno,\total,\commutative,\associative)$}-{OWF}s}
\label{sec:nonstrong-cases}

It remains to prove the $27$ ``nonstrong'' cases.  All $27$ have 
$\p \neq \np$ as a necessary condition.  For each of them, we also completely
characterize the existence of such {\OWF}s by $\p \neq \np$.  

First, we consider two ``total'' and
``nonstrong'' cases in Lemmas~\ref{lem:nyyn} and~\ref{lem:nynn} below.
Note that Hemaspaandra, Pasanen, and
Rothe~\cite{hem-pas-rot:j:strong-noninvertibility} constructed one-way
functions that in fact are not strongly noninvertible.  Unlike
Lemmas~\ref{lem:nyyn} and~\ref{lem:nynn}, however, they did not
consider associativity and commutativity.  Note that, in the proofs of
Lemmas~\ref{lem:nyyn} and~\ref{lem:nynn}, we achieve
``nonstrongness'' while ensuring that the functions constructed are
s-honest.  That is, they are not ``nonstrong'' because they are not
s-honest, but rather they are ``nonstrong'' because they are not strongly
noninvertible.

\begin{lemma}
\label{lem:nyyn}
If $\p \neq \np$ then there exist $(\no,\yes,\yes,\no)$-{\OWF}s.
\end{lemma}

\begin{proofs}
Assuming $\p \neq \np$, we define an $(\no,\yes,\yes,\no)$-{\OWF} that
is akin to a function constructed 
in Theorem~3.4
of~\cite{hem-pas-rot:j:strong-noninvertibility}
(which is currently most easily 
available via Theorem~3
of~\cite{hem-pas-rot:c:strong-noninvertibility}).

Define a function $\sigma : \sigmastar \times \sigmastar
\rightarrow \sigmastar$ by 
\[
\sigma(x,y) = \left\{
\begin{array}{ll}
1\rho(x)                & \mbox{ if $x = y$} \\ 
0\min(x,y)\max(x,y) & \mbox{ if $x \neq y$,}
\end{array}
\right.
\]
where $\min(x,y)$ denotes the lexicographically smaller of $x$
and~$y$, $\max(x,y)$ denotes the lexicographically greater of $x$
and~$y$, and $\rho : \sigmastar \rightarrow \sigmastar$ is a total
one-ary one-way function, which exists assuming $\p \neq \np$.  Note that
$\sigma$ is polynomial-time computable, total, honest, and s-honest.
Clearly, if $\sigma$ could be inverted in polynomial time then $\rho$
could be too.  Thus, $\sigma$ is a one-way function.  However,
although $\sigma$ is s-honest, it is not strong.
To prove that $\sigma$ is not strongly noninvertible, we show that it
is invertible with respect to each of its arguments.  
Define a
function $f_1 : \sigmastar \rightarrow \sigmastar$ by
\[
f_1(a) = \left\{
\begin{array}{ll}
y & \mbox{ if $(\exists x, y, z \in \sigmastar)\, [a = \pair{x,0z} \land
             z = xy \land x <_{\rm lex} y]$} \\
y & \mbox{ if $(\exists x, y, z \in \sigmastar)\, [a = \pair{x,0z} \land
             z = yx \land y <_{\rm lex} x]$} \\
x & \mbox{ if $(\exists x, z \in \sigmastar)\, [a = \pair{x,1z}]$} \\
\varepsilon & \mbox{ otherwise,}
\end{array}
\right.
\]
where $x <_{\rm lex} y$ indicates that $x$ is strictly smaller than
$y$ in the lexicographic ordering of~$\sigmastar$.  Note that $f_1$ is
in~$\fp$ and that $f_1$ inverts $\sigma$ with respect to the first
argument.  Although this is already enough to defy strong
noninvertibility of~$\sigma$, we note that one can analogously show
that $\sigma$ also is invertible with respect to the second argument.

To see that $\sigma$ is commutative, note that
if $x \neq y$ then $\sigma(x,y) = 0\min(x,y)\max(x,y) = \sigma(y,x)$.
(Although the $x=y$ case does not need to be discussed to 
establish commutativity, for completeness we mention that 
if $x = y$ then $\sigma(x,y) = 1\rho(x) = \sigma(y,x)$.)
To see that $\sigma$ is nonassociative, note that
$\sigma(\sigma(1,0),001) = \sigma(001,001) = 1\rho(001) \neq 0100001 =
\sigma(1,00001) = \sigma(1,\sigma(0,001))$.

Thus, $\sigma$ is an $(\no,\yes,\yes,\no)$-{\OWF}, which completes the proof.
\end{proofs}

\begin{lemma}
\label{lem:nynn}
If $\p \neq \np$ then there exist $(\no,\yes,\no,\no)$-{\OWF}s.
\end{lemma}

\begin{proofs}
Assume that $\p \neq \np$.  So there exists a total one-argument
one-way function $\rho : \sigmastar \rightarrow \sigmastar$.  
In Theorem~3.4
of~\cite{hem-pas-rot:j:strong-noninvertibility}, a function
$\sigma : \sigmastar \times \sigmastar \rightarrow \sigmastar$ is
constructed as follows:
\[
\sigma(x,y) = \left\{
\begin{array}{ll}
1\rho(x) & \mbox{ if $x = y$} \\ 
0xy      & \mbox{ if $x \neq y$.}
\end{array}
\right.
\]
It is shown
in~\cite{hem-pas-rot:j:strong-noninvertibility} that $\sigma$ is a
total, s-honest one-way function that is not strongly noninvertible.

To see that $\sigma$ is noncommutative, note that
\[
\sigma(0,1) = 001 \neq 010 = \sigma(1,0) .
\]
To see that $\sigma$ is nonassociative, note that
\[
\sigma(\sigma(0,1),001) = 1\rho(001) \neq 0001001 = \sigma(0,\sigma(1,001)) .
\]
Thus, $\sigma$ is an $(\no,\yes,\no,\no)$-{\OWF}, which completes the proof.
\end{proofs}

Next, we observe that the two remaining ``total'' and ``nonstrong''
cases are connected: Lemma~\ref{lem:nyyy-implies-nyny} shows that,
given an $(\no,\yes,\yes,\yes)$-{\OWF}, one can construct an
$(\no,\yes,\no,\yes)$-{\OWF}.  Thus, by
Lemma~\ref{lem:eliminating-totality}, characterizing via $\p \neq \np$ 
just the case of
$(\no,\yes,\yes,\yes)$-{\OWF}s will suffice to solve all the
four remaining cases (namely, NYYY, NYNY, NNYY, and NNNY) 
at once.

\begin{lemma}
\label{lem:nyyy-implies-nyny}
If there exist $(\no,\yes,\yes,\yes)$-{\OWF}s, then there exist
$(\no,\yes,\no,\yes)$-{\OWF}s.
\end{lemma}

\begin{proofs}
The proof uses the construction presented in Lemma~\ref{lem:yyny},
except that we now start
from an $(\no,\yes,\yes,\yes)$-{\OWF} $\sigma$ instead of a
$(\yes,\yes,\yes,\yes)$-{\OWF} as in Lemma~\ref{lem:yyny}.
Constructing $\rho$ from $\sigma$ according to the proof of
Lemma~\ref{lem:yyny} yields a total, noncommutative, associative
one-way function that is not strongly noninvertible.
\end{proofs}

We now turn to completely characterizing the existence of
$(\no,\yes,\yes,\yes)$-{\OWF}s.
A transformation from the literature that might seem 
to come close to establishing ``if $\pisnotnp$, then
$(\no,\yes,\yes,\yes)$-{\OWF}s  exist'' has been shown to be 
flawed unless an unlikely complexity class collapse 
occurs.\footnote{In more detail:  Rabi and 
Sherman~\cite{rab-she:t-no-URL:aowf,rab-she:j:aowf}, assuming $\p \neq \np$, 
constructed a {\em nontotal}, commutative, associative 
(in a slightly weaker model of associativity for partial 
functions that completely 
coincides with our model when speaking of total functions) one-way
function that appears to fail to possess strong noninvertibility.  They also
proposed a construction that they claim can be used to transform every
nontotal AOWF whose domain is in $\p$ to a total AOWF.  However, their
claim does not provide an $(\no,\yes,\yes,\yes)$-{\OWF},
due to some
subtle technical points.
First, Rabi and Sherman's construction---even if their claim were
valid---is not applicable to the nonstrong, nontotal, commutative
AOWF they construct, since this function seems to not have a domain
in~$\p$.  Second, it it is not at all clear that their 
above-mentioned ``construction to add totality''
has the properties they assert 
for it.  In particular, let
$\up$ as usual denote Valiant's~\cite{val:j:checking} class
representing
``unambiguous polynomial time.'' 
Hemaspaandra and
Rothe showed in~\cite{hem-rot:j:aowf} that any proof that the
Rabi--Sherman claim about their transformation's 
action is in general 
valid would immediately prove that $\up = \np$, which is considered
unlikely.  
} 
However, the following result of Rabi and Sherman does 
provide 
evidence that $(\no,\yes,\yes,\yes)$-{\OWF}s
indeed exist.  

\begin{theorem} 
\label{thm:factoring-not-in-p-implies-nyyy}
{\bf \cite{rab-she:j:aowf,rab-she:t-no-URL:aowf}}~~If
factoring is not in polynomial time, then
there 
exist $(\no,\yes,\yes,\yes)$-{\OWF}s.
\end{theorem}

We now improve that sufficient condition to $\p \neq \np$.

\begin{lemma}
\label{lem:foo-new}
If $\p \neq \np$ then there exist $(\no,\yes,\yes,\yes)$-{\OWF}s
and $(\no,\yes,\no,\yes)$-{\OWF}s.
\end{lemma}

\begin{proofs}
By Lemma~\ref{lem:nyyy-implies-nyny}, it suffices to handle the 
case of 
$(\no,\yes,\yes,\yes)$-{\OWF}s.  So, assume $\p \neq \np$.  
This implies that 
there exists a total, one-way function $f: \sigmastar \rightarrow 
\sigmastar$.  
Define the
function $g : \sigmastar \times \sigmastar \rightarrow \sigmastar$ by
\[
g(x,y) = \left\{
\begin{array}{ll}
0f(a) & \mbox{ if $x=1a$ and $y=1a$} \\
\varepsilon & \mbox{ otherwise.}
\end{array}
\right.
\]
$g$ is clearly a one-way function.  $g$ also is 
clearly total and commutative.  $g$ is associative since
it is not hard to see that 
$(\forall a,b,c)[ g(a, g(b,c)) = g(g(a,b),c) = \varepsilon]$.
Though $g$ is easily seen to be s-honest, $g$ fails
to be strongly noninvertible, and so is not strong.
In particular, given the output and a purported first argument,
here is how to find a second argument consistent with the 
first argument when one exists.  
If the output is 
$\varepsilon$ and the purported first argument is $z$, 
then output $\varepsilon$ as a second argument. 
If the output is 
$0y$ and the purported first argument is $1x$, 
then if $f(x)=y$ a good second argument 
is $1x$.  In every other case, the output and 
purported first argument cannot have any second argument
that is consistent with them, so we safely (though irrelevantly,
except for achieving totality of our inverter if one desires
that) in this 
case  have our inverter output $\varepsilon$.
\end{proofs}

\begin{theorem}
\label{thm:nonstrong-solved-cases}
For each $\total, \commutative, \associative \in \{\yes, \no,
\yesno\}$,
there exist $(\no, \total, \commutative, \associative)$-{\OWF}s if and
only if $\p \neq \np$.
\end{theorem}

The proof of Theorem~\ref{thm:nonstrong-solved-cases} follows
immediately from Lemmas~\ref{lem:nyyn}, \ref{lem:nynn},
and~\ref{lem:foo-new}, via
Lemmas~\ref{lem:eliminating-yesno},
\ref{lem:eliminating-one-direction},
and~\ref{lem:eliminating-totality}.  

In conclusion, this paper studied the question of whether one-way
functions can exist, where one imposes either possession,
nonpossession, or being oblivious to possession of the properties of
strongness, totality, commutativity, and associativity.  We have shown
that $\p \neq \np$ is a necessary and sufficient condition in each
of the possible $81$ cases.  

\bibliographystyle{alpha}

{

} 

\end{document}